\pdfoutput=1
\documentclass[reprint,floatfix,twocolumn,superscriptaddress,aps]{revtex4-2}
\usepackage{graphicx}
\usepackage{amsmath}
\usepackage{amsfonts}
\usepackage{amssymb}
\usepackage{xcolor}
\usepackage{bm}
\usepackage[utf8]{inputenc}
\usepackage[T1]{fontenc}
\usepackage{lmodern}
\usepackage[hidelinks]{hyperref}

\newcommand{\av}[1]{\left\langle #1 \right\rangle}
\newcommand{\x}{\mathbf{x}}
\newcommand{\D}{\mathbf{D}}
\newcommand{\Dbar}{\bar{\mathbf{D}}}
\newcommand{\Dhat}{\hat{\mathbf{D}}}

\begin{document}

\title{And yet it moves -- or does it? Spurious drifts at equilibrium}

\author{Pierre Ronceray}\email{pierre.ronceray@univ-amu.fr}
\affiliation{Aix-Marseille Universit\'e, CNRS, CINaM, Turing Centre for Living Systems, Marseille, France}

\begin{abstract}
  When diffusion depends on the state of a system, the drift measured
  from its trajectories contains a spurious part unrelated to forces. At equilibrium, particles can then appear to sediment, repel each
  other, even rotate -- the same way whether the movie is played
  forward or backward. Drifts are conditional averages, not
  currents: they carry meaning only together with a consistently
  inferred diffusion tensor. Paired with the wrong one, they make
  passive particles look active. I show how
  to infer spurious drifts from trajectory moments alone, enabling
  thermodynamically consistent inference.
\end{abstract}

\maketitle

Diffusion is rarely homogeneous in Brownian systems. Colloids diffuse
more slowly near walls and near each
other~\cite{brenner1961,lancon2001,dufresne2000}, proteins
anisotropically in crowded membranes, tracers in liquid crystals along
the director. The diffusion tensor $\D(\x)$ then depends on the state
of the system, the thermal noise is multiplicative, and the drift of
the Langevin equation contains a convention-dependent \emph{spurious}
term, also called noise-induced drift or spurious
force~\cite{vankampen1981,hanggi1978,klimontovich1990,lau2007,volpe2016,serov2020}.
This is textbook material~\cite{gardiner2009}.

It has become a practical problem because drifts are now
\emph{measured}. Particle tracking routinely turns trajectories into
drift maps, which are read as force
fields~\cite{sainis2007,sarfati2017,frishman2020,bruckner2020}, tested
for broken detailed balance~\cite{battle2016,gnesotto2018,li2019}, or
used to predict transport~\cite{schnitzer1993}. Read in these ways, drift maps produce illusions. In four equilibrium
systems, particles appear to sediment in a channel where the density
stays flat; to circulate around a liquid-crystal defect, or to be
propelled through its core; and, in published experimental data, to
attract each other four times more weakly than they do, or to sit above
a wall in a potential off by $k_{\rm B}T$. The sharpest test is to play
the movie backwards: the drift map of an equilibrium trajectory is
unchanged, and rotates the same way. A drift that survives time
reversal, I argue, is not a motion. The mechanism is always the same:
conditioning on a position selects an ensemble that is not at
equilibrium, and the drift measures how that ensemble relaxes. At
equilibrium, the particles found at a given place are, on average, on
their way back to where they came from.

The ingredients are known --- the drift is not Nelson's current
velocity~\cite{nelson1967}, mid-point averages measure
currents~\cite{frishman2020}, force inference can correct for a scalar
$\nabla D$~\cite{serov2020}, drift and Boltzmann force measurements
disagree~\cite{brettschneider2011,lavaud2021} --- but their consequences
are poorly appreciated: a correctly measured drift, paired with the
wrong diffusion tensor, makes a model that circulates and dissipates
although fitted to equilibrium data. To avoid this, I introduce a
thermodynamically consistent inference method, in which the spurious
drift is measured from trajectory moments alone, without differentiating
an inferred diffusion field.

\paragraph*{Multiplicative noise and spurious drifts.}
Consider an overdamped particle with diffusion tensor $\D(\x)$ and
potential $V(\x)$ in contact with a thermal bath ($k_{\rm B}T=1$).
Equilibrium requires the Boltzmann distribution $P\propto e^{-V}$
with zero probability current, which fixes the It\^o
dynamics~\cite{vankampen1981,lau2007}
\begin{equation}
  \label{eq:langevin_equilibrium}
  \mathrm{d}\x_t = \Phi\,\mathrm{d}t + \sqrt{2\D(\mathbf{x}_t)}\cdot\mathrm{d}\mathbf{W}_t ,
  \quad
  \Phi \equiv \nabla\cdot\D - \D\nabla V ,
\end{equation}
with $[\nabla\cdot\D]_\nu = \sum_\mu \partial_\mu D_{\mu\nu}$. The term
$\nabla\cdot\D$ is the \emph{spurious drift}. It is not optional:
without it the stationary state is not Boltzmann's. For a scalar
diffusivity, evaluating the noise at the end point of each step (the
H\"anggi--Klimontovich, or anti-It\^o,
convention~\cite{hanggi1978,klimontovich1990}) hides it from the
equation, and $\nabla D$ is a gradient, which can be
dressed as an effective potential. Neither holds for a tensor: the
end-point correction then depends on which square root of $2\D$
multiplies the noise~\cite{lau2007}, and does not generally cancel out in the anti-It\^o convention (Appendix~B); and a
tensor divergence is not a gradient and cannot be
absorbed into an effective potential, entropic or otherwise. Which convention a system
realizes out of equilibrium~\cite{volpe2016,bo2017}, and how
Eq.~\eqref{eq:langevin_equilibrium} emerges from the frictional limit
of an inertial system~\cite{sancho1982,kupferman2004,hottovy2015}, are
separate questions not tackled here.

Operationally, a drift is measured by averaging the increments
$\Delta\x = \x_{t+\Delta t}-\x_t$ of the steps that visit a given
location $\x$. Visit it \emph{when}? Conditioning on the start, the
middle or the end of the step gives three different fields, which as
$\Delta t\to 0$ tend to
\begin{subequations}
  \label{eq:drifts}
  \begin{align}
    \hat{v}_{\rm pre}(\x) &\equiv
    \av{ \tfrac{\Delta\x}{\Delta t} \,\big|\, \x_t = \x }
    = \Phi , \label{eq:pre}\\
    \hat{v}_{\rm mid}(\x) &\equiv
    \av{ \tfrac{\Delta\x}{\Delta t} \,\big|\, \tfrac{\x_t+\x_{t+\Delta t}}{2} = \x }
    = \mathbf{J}/P , \label{eq:mid}\\
    \hat{v}_{\rm post}(\x) &\equiv
    \av{ \tfrac{\Delta\x}{\Delta t} \,\big|\, \x_{t+\Delta t} = \x }
    = \Phi - 2\nabla\cdot(\D P)/P , \label{eq:post}
  \end{align}
\end{subequations}
with $\mathbf{J} = \Phi P - \nabla\cdot(\D P)$ the probability current
(Appendix~B). Drift-based inference measures the pre-point drift;
the mid-point velocity is the current per particle; the post-point
drift is Nelson's backward velocity~\cite{nelson1967}.

The three behave differently when the trajectory is time-reversed:
every increment changes sign, the start of a step becomes its end,
and the middle stays the middle. For any stationary process the
reversed movie has pre-point drift $-\hat v_{\rm post}$ and mid-point
velocity $-\hat v_{\rm mid}$: the mid-point velocity is odd under time
reversal, as a velocity should be, whereas the pre-point drift alone
has no symmetry. At equilibrium, the reversed movie is statistically
indistinguishable from the original. Its mid-point velocity must
equal both $\hat v_{\rm mid}$ and $-\hat v_{\rm mid}$: it vanishes, at
any $\Delta t$. Its pre-point drift must equal $\Phi$, so that $\hat
v_{\rm post} = -\Phi$. At equilibrium, the drift is even under time reversal:
it cannot be a velocity.

\begin{figure}
  \includegraphics[width=\columnwidth]{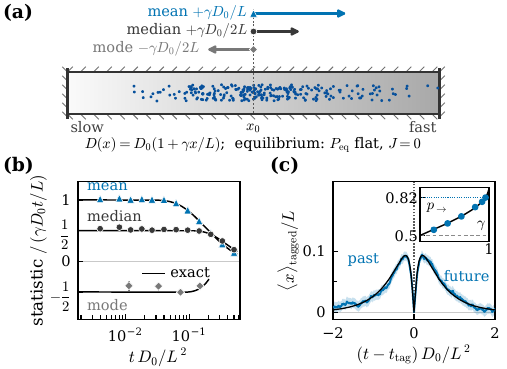}
  \caption{\label{fig:channel}%
    \textbf{Apparent sedimentation without flux.}
    (a)~Zero potential, reflecting walls at $x=\pm L$ and
    $D(x) = D_0(1+\gamma x/L)$ (shading) with $\gamma = 0.95$, flat
    $P_{\rm eq}$. Dots: an ink drop released at $x_0=0$, at $t =
    0.06\,L^2/D_0$; arrows: velocities of its mean, median and mode.
    (b)~Mean ($\blacktriangle$), median ($\bullet$) and mode
    ($\blacklozenge$) of the drop divided by $\gamma D_0t/L$: plateaus
    $1$, $1/2$ and $-1/2$ until the walls are felt.
    (c)~Mean position of the particles found in $|x|<0.1L$ at
    $t_{\rm tag}$ in the stationary state ($\pm2$~SEM). Inset:
    probability $p_\rightarrow$ of first exit through the fast end,
    Eq.~\eqref{eq:splitting}. Symbols and bands: Monte Carlo; black
    lines: exact.}
\end{figure}

\paragraph*{Apparent sedimentation in a channel.}
Consider a particle in a segment $|x|<L$ with hard walls, no
potential, and $D(x) = D_0(1+\gamma x/L)$, $\gamma<1$
(Fig.~\ref{fig:channel}a): a minimal model of colloids between
near-parallel walls, where the spurious drift was first
observed~\cite{lancon2001}, and the classical ``paradox'' of
diffusion in a diffusivity
gradient~\cite{schnitzer1993,vanmilligen2005}. The stationary density
is flat. Yet an ensemble released at the center, an ink drop, moves
on average towards the fast side, $\av{x_t} = \gamma D_0 t/L$, since
$\mathrm{d}\av{x}/\mathrm{d}t = \av{D'}$ (Fig.~\ref{fig:channel}a,b).
Do the particles sediment?

They do not: the drop skews rather than translates. Its median moves
half as fast as its mean, and its mode moves the other way, at
$-\gamma D_0/2L$, because the short-time propagator has a long tail on
the fast side and a peak displaced towards the slow side
(Fig.~\ref{fig:channel}b). Where the drop ``goes'' depends on the
statistic, and none is a current. Time-reversal symmetry is exposed by
tagging the particles found near the center at $t_{\rm tag}$ in the
stationary state and following them forward and backward in time:
their mean position is symmetric in $t-t_{\rm tag}$
(Fig.~\ref{fig:channel}c).  They move towards the fast side after
$t_{\rm tag}$ because they came from it. A three-state chain with a
fast and a slow bond captures the essence of this paradox
(Appendix~A). Exits are asymmetric too: $82\%$ of the particles released at
the center first leave through the fast end, yet reservoirs at equal
concentrations attached to the ends exchange no net flux, because
$82\%$ of the particles first \emph{arriving} at the center also come
from the fast end (Fig.~\ref{fig:channel}c, inset; Appendix~A).

\begin{figure*}
  \includegraphics[width=\textwidth]{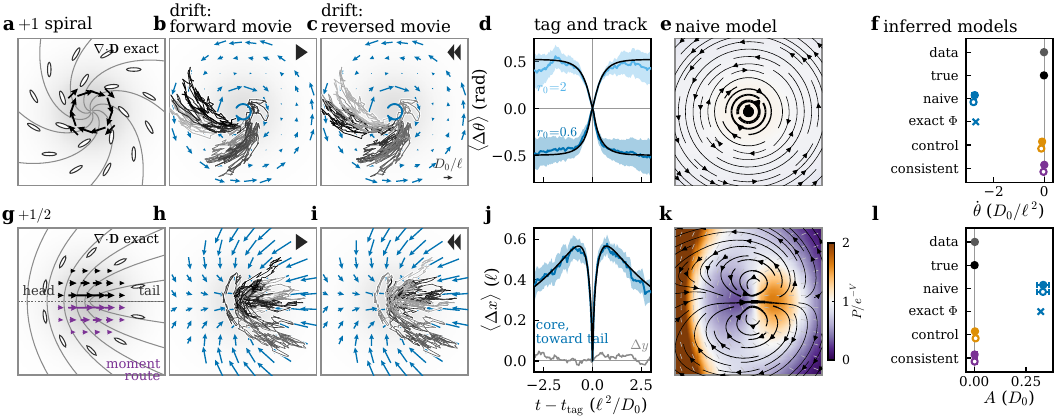}
  \caption{\label{fig:vortex}%
    \textbf{Apparent rotation and propulsion around topological defects, and what inference makes of them.}
    Tracer near a $+1$ spiral (top) and a $+1/2$ defect (bottom) of a nematic, in the trap $V=r^2/2\ell^2$
    (shading: $e^{-V}$), with $\D$ of Eq.~\eqref{eq:tensorD}, $a=0.9$, $\xi=\ell/4$. (a,g)~Diffusion ellipses, director lines and spurious drift $\nabla\cdot\D$ (exact; lower half of g:
    moment-route estimate). (b,c,h,i)~Binned pre-point drift of the movie and of its reversal (azimuthal part in
    b,c), with one trajectory ($10\,\ell^2/D_0$; dark = late). (d,j)~Mean angular (d) or axial (j) displacement of
    the particles found at $r_0$ (d) or in the core, $r<0.15\,\ell$ (j), at $t_{\rm tag}$; black: exact (Appendix~C). (e,k)~Stationary density relative to $e^{-V}$ and current (width $\propto|\mathbf J|$) of the naive
    model. (f,l)~Mean winding rate (f) or lobe circulation $A=\av{\mathrm{sgn}(y)\,\x\times\dot\x}/2$ (l) of the data
    and of the models inferred from the movie ($\bullet$) or its reversal ($\circ$); $\times$: naive model with the exact $\Phi$. Bands and bars: $\pm2$~SEM
    (data, true model) or $\pm2$ bootstrap s.d. (inferred models).}
\end{figure*}

\paragraph*{Apparent rotation and propulsion around defects.}
A tensor divergence can circulate when the principal axes of $\D$
rotate in space~\cite{farago2017}. A natural setting is a tracer in a
nematic liquid crystal, which diffuses faster along the director
$\hat n$~\cite{loudet2004}. To leading order in the order parameter
$s$, symmetry imposes
\begin{equation}
  \label{eq:tensorD}
  \D = D_0\,\openone + \beta\,\mathbf Q ,\qquad
  \mathbf Q = s\,\big(\hat n\hat n^{\sf T} - \tfrac12\openone\big) ,
\end{equation}
with eigenvalues $D_0(1\pm a s)$, $a=\beta/2D_0<1$. Around a defect
of charge $k$, $\hat n$ makes the angle $k\theta+\phi_0$ with the $x$
axis, the core is melted, $s(r)\to0$ as $r\to0$, and, in complex
notation,
\begin{equation}
  \label{eq:divQ}
  \nabla\cdot\D = aD_0\Big(s' + \frac{2ks}{r}\Big)\,e^{i[(2k-1)\theta + 2\phi_0]} .
\end{equation}
The spurious drift winds $2k-1$ times around the defect and decays as
$1/r$: asters and vortices ($k=1$, $\phi_0=0$ or $\pi/2$) produce an
apparent radial repulsion or attraction, $+1$ spirals an apparent
rotation, and a $+1/2$ defect an apparent push through its core, from
head to tail. For illustration purposes, I consider a tracer held near the
defect by a harmonic trap of length $\ell$, $V=r^2/2\ell^2$ (optical
tweezers at a photopatterned defect~\cite{koizumi2020}, or core
attraction~\cite{wang2016}), a
melted core of size $\xi=\ell/4$, and $a=0.9$. The example is
idealized: the anisotropy far exceeds that of colloids in
nematics~\cite{loudet2004}, and elastic forces are ignored (Appendix~E, Sec.~S6).

Around a spiral whose fast axis is tilted by $45^\circ$ from the
radius, the spurious drift is a clockwise
vortex  (Fig.~\ref{fig:vortex}a). The pre-point drift measured on equilibrium trajectories
(Fig.~\ref{fig:vortex}b) also circulates: read as a velocity, it
predicts over $80$ clockwise turns per recording, whereas on average the
tracers make none (Fig.~\ref{fig:vortex}f; Appendix~C). Played backwards, the
movie has the same drift map, turning the same way
(Fig.~\ref{fig:vortex}c): a rotation that survives time reversal is not
a rotation.

The conditioned
rotation is a gearing effect: the tilted fast axis couples radial and
angular motion like a screw thread, so that a particle moving outwards
turns clockwise and one moving inwards counterclockwise (Appendix~A,~C).  Particles tagged at a radius $r_0$ form an ensemble out of
radial equilibrium; as it relaxes it turns by a finite angle,
clockwise for shells tagged inside, counterclockwise outside, and
identically backward in time (Fig.~\ref{fig:vortex}d). The drift map
must indeed counter-rotate at large $r$, since at equilibrium the
density-weighted torque of the drift vanishes for any $\D(\x)$ without
boundaries, $\av{\x\times\Phi} = 0$ (Appendix~C). An equilibrium
drift map can thus swirl locally, not rotate as a whole. The $+1/2$ defect
does not rotate, but its drift pushes tracers through the core towards
the tail (Fig.~\ref{fig:vortex}h,i), in both movies: the particles
found in the core move towards the tail because they came from it
(Fig.~\ref{fig:vortex}j), the splay tail being the fast channel into
the core, like the fast end of the channel in Fig.~\ref{fig:channel}.

\paragraph*{Consistent inference: the moment route.}
Drift statistics alone cannot tell a spurious drift from a force:
whatever the conditioning, Eqs.~\eqref{eq:drifts} return combinations
of $\Phi$ and $\nabla\cdot(\D P)/P$, which do not fix $\nabla\cdot\D$
(Appendix~B). A measured drift is therefore meaningful only together with
the diffusion tensor. The \emph{naive} reading, which pairs
$\hat\Phi$ with a constant $\Dbar$, attributes the whole drift to a
force, $\Dbar^{-1}\hat\Phi$. The missing information is in
the quadratic variation, which measures
$\D$~\cite{vestergaard2014,frishman2020}; fitting $\D(\x)$ and
differentiating it, however, amplifies noise and fitting error. A
more robust route follows from what a force does: it
either drives a current or holds a density gradient against diffusion,
$\D\mathbf F = \mathbf J/P + \D\nabla\ln P$, in or out of
equilibrium. Subtracting this from the drift
$\Phi = \nabla\cdot\D + \D\mathbf F$ leaves the spurious drift as a
sum of three measurable moments,
\begin{equation}
  \label{eq:identity}
  \nabla\cdot\D = \hat v_{\rm pre} - \hat v_{\rm mid} - \D\nabla\ln P :
\end{equation}
the pre-point drift, the mid-point velocity, and the quadratic
variation contracted with the score $\nabla\ln P$, itself estimated
from positions by score matching~\cite{hyvarinen2005,frishman2020}. No
diffusion field is differentiated. At equilibrium the mid-point term
vanishes and Eq.~\eqref{eq:identity} is Boltzmann inversion weighted
by $\D$; out of equilibrium, it carries the current. I call this the
\emph{moment route} to spurious drift inference (Appendix~F; implemented in
SFI~\cite{frishman2020,sfi_zenodo}). It splits the measured
drift into a force part $\D\hat{\mathbf F}$ and a spurious drift
$\hat\sigma$, whose relative size
$\nu = \|\hat\sigma\|/\|\hat\Phi\|$ tells how much of the drift is
not a force (Appendix~F).

Suppose now that an analyst fits a model to the defect trajectories. With
which diffusion tensor should the measured drift $\hat\Phi$ be paired?
The naive choice, the average $\Dbar$, realizes the
drift-as-velocity reading: around the spiral the model winds at the rate
the drift map suggests, and dissipates; around the $+1/2$ it pumps
tracers through the core and back along the flanks
(Fig.~\ref{fig:vortex}e,k). Fitted to the reversed movies, it does the
same (Fig.~\ref{fig:vortex}f,l). Nothing was mismeasured --- $\hat\Phi$
and $\Dbar$ are both accurate~\cite{frishman2020} --- the error lies in
the pairing, which turns an inert spurious drift into a force. The
currents this force drives mimic those of a dry active nematic with
activity $\zeta=-\beta$, since $\beta\nabla\cdot\mathbf Q$ is the active
force density~\cite{ramaswamy2003} (Appendix~C): a naive analysis makes a
passive tracer look active. Scalar-diffusivity inference would not help,
as ${\rm Tr}\,\D$ is uniform here~\cite{serov2020} (Sec.~S7). Pairing
$\hat\Phi$ with a field $\Dhat(\x)$ fitted independently to the
quadratic variation, the \emph{control} model, is exact in principle but
still winds slowly: nothing ties the two fits together (Appendix~E). The
\emph{consistent} model instead rebuilds its drift as $\nabla\cdot\Dhat
+ \Dhat\hat{\mathbf F}$, with the moment-route force $\hat{\mathbf F}$
of Eq.~\eqref{eq:identity}. It alone neither winds nor circulates, from
either movie (Fig.~\ref{fig:vortex}f,l), even at weak anisotropy and
large localization noise (Appendix~E). Out of equilibrium the naive reading
fails the other way: tracers advected around the spiral by a rigid
rotation tuned to cancel the apparent one keep the Boltzmann density yet
wind at $2.7$~rad per $\ell^2/D_0$; the naive model winds at $0.3$, the consistent one at $2.6$ (Sec.~S5). Naive inference sets a passive tracer
spinning and nearly stops a spinning one: a model whose currents differ
from those of its data has missed a spurious drift.

\begin{figure}
  \includegraphics[width=\columnwidth]{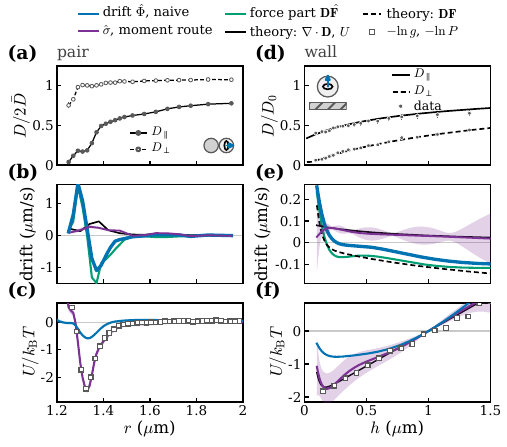}
  \caption{\label{fig:pair}%
    \textbf{Hydrodynamic interactions masquerading as forces.}
    Left column: colloid pairs in a depletion monolayer~\cite{king2025}, separation $r$. Right column: a colloid above a
    wall~\cite{lavaud2021}, gap $h$.
    (a,d)~Measured diffusivities (dots) along ($D_\parallel$) and normal to ($D_\perp$) the line of centers
    or the wall; (a) relative coordinate, in units of $2\bar D$; lines in (d): standard near-wall approximations (Fax\'en,
    Brenner~\cite{brenner1961}), as in Ref.~\cite{lavaud2021}. Sketches: squeeze flow flattens the diffusion
    ellipse, and the It\^o drift (arrow) points away.
    (b,e)~Measured drift, split by the moment route into spurious drift and force part, against
    Eq.~\eqref{eq:pairforce} on the measured diffusivities (b; $40$-nm bins, shape unresolved in the
    well, Appendix~D) and the exact $\nabla\cdot\D$ and $\D\mathbf F$ (e).
    (c,f)~Potentials inferred from the same drift with a constant $\Dbar$ (naive) and by the moment route;
    Boltzmann inversion, $-\ln g(r)$ and $-\ln P(h)$; true potential (f). Bands: $\pm2$~SEM over
    $61$ videos (pair), $\pm2$ block SE (wall).}
\end{figure}

\paragraph*{Hydrodynamic interactions masquerading as forces in colloidal experiments.}
Multiplicative noise is the rule for interacting colloids:
hydrodynamic coupling makes the mobility of each particle depend on
the positions of the others. For the separation $\mathbf{r}$ of
two identical spheres, symmetry imposes
$\D_{\rm rel} = D_\parallel(r)\, \hat{r}\hat{r}^{\sf T} +
D_\perp(r)(\openone - \hat{r}\hat{r}^{\sf T})$, with $D_\parallel \to
0$ at contact, where the fluid must be squeezed out of the gap, and
$D_\parallel$ rising with $r$. The spurious drift is radial,
\begin{equation}
  \label{eq:pairforce}
  \left[\nabla\cdot\D_{\rm rel}\right]_r = D_\parallel'(r)
  + (d-1)\,\frac{D_\parallel(r) - D_\perp(r)}{r}
\end{equation}
in $d$ dimensions (Appendix~D), and points outwards near contact
because the $D_\parallel'$ term dominates there, so that it masks
attractions. In a depletion-attracted monolayer of $2300$ spheres of
diameter $1.3\,\mu$m, imaged every $7$~ms~\cite{king2025},
$D_\parallel$ drops nearly fivefold at contact, and the measured drift
contains an outward spurious drift matching Eq.~\eqref{eq:pairforce}
in sign and magnitude (Fig.~\ref{fig:pair}a,b). Read with a constant
diffusion coefficient, it yields a well of $-0.6\,k_{\rm B}T$: slowed bond motion
looks like a weak force, and the spurious drift pushes contacts apart.
The moment route returns $-2.5\,k_{\rm B}T$, four times deeper, as does $-\ln g(r)$
(Fig.~\ref{fig:pair}c), unchanged when the frame interval grows
$64$-fold, whereas the naive well vanishes (Appendix~D).
Hydrodynamic interactions modulate where the pair diffuses fast,
which biases conditional displacements; they exert no force and
create no structure. The far-field mobility tensors of Oseen
and Rotne--Prager are divergence-free in
3D~\cite{rotne1969,ermak1978}, but not for pairs confined to a plane,
the geometry of drift-based pair
measurements~\cite{dufresne2000,sainis2007,sarfati2017}: there the
naive repulsion is long-ranged, $\simeq(3R/4r)\,k_{\rm B}T$ for spheres of radius $R$~\cite{bleibel2014} (Appendix~D).

The same hydrodynamics acts between a colloid and a wall, where the physical force and the state-dependent diffusion were recently measured directly~\cite{lavaud2021}. Figure~\ref{fig:pair}d--f
reanalyzes an experimental trajectory of Ref.~\cite{lavaud2021}: a sphere of radius
$1.5\,\mu$m above a glass plate, tracked at $60$~Hz for $32$~min, its
gap $h$ to the wall resolved to $10$~nm. Squeeze flow slows the
vertical diffusion, $D_\perp \propto h$ at contact~\cite{brenner1961};
gravity pulls down over $\ell_B = 0.47\,\mu$m, and a screened
electrostatic repulsion pushes up within $30$~nm. The moment route
splits the measured drift (Fig.~\ref{fig:pair}e): its spurious part
follows Brenner's $D_\perp'(h)$ with no adjustable parameter, its
force part gravity plus electrostatics. The spurious drift dominates
($\nu = 0.74\pm0.15$): at the most probable height, where the force
vanishes, the particle drifts away from the wall at $D_\perp' =
4\,\mu$m/min, whichever way the movie is played (Sec.~S10). Read with
a constant diffusion coefficient, the drift yields a potential off by
$1\,k_{\rm B}T$ near the wall (Fig.~\ref{fig:pair}f); the moment-route
potential agrees with gravity plus electrostatics, and with the Boltzmann
inversion of the height histogram performed in Ref.~\cite{lavaud2021}, to $0.2\,k_{\rm B}T$ (Sec.~S10).

\paragraph*{Discussion.}
Drifts are not currents. Currents, such as the mid-point velocity,
the winding of a trajectory or the entropy
production~\cite{roldan2010,battle2016,gnesotto2018,li2019}, are odd
under time reversal, vanish at equilibrium, and signal actual motion.
Drifts are conditional averages: their time-even part
$\nabla\cdot(\D P)/P$, the whole drift at equilibrium, encodes how a
conditioned ensemble relaxes; only their time-odd part $\mathbf J/P$
is a current (Appendix~B). A curl of
the drift does not signal nonequilibrium: already for a constant
anisotropic $\D$, the potential condition bears on
$\D^{-1}(\Phi-\nabla\cdot\D)$, not on $\Phi$~\cite{gardiner2009},
whereas the density-weighted torque of the drift, equal to that of
the current, is a valid test without boundaries (Appendix~C). More
insidious is any model that pairs the drift with a diffusion tensor
inconsistent with it: it circulates, dissipates or repels, although
fitted to equilibrium data.

The pitfall is not hypothetical. Simulations that omit the spurious
drift, and drifts read as transport, have repeatedly drawn
published corrections and critiques, in oceanography~\cite{holloway1994,visser1997}
and in cell biophysics, for
the heterogeneity of intracellular viscosity~\cite{skora2024} and a
proposed accumulation of particles in low-diffusivity
regions~\cite{elife2024}.  Nor has the controversy about noise-induced
drifts~\cite{volpe2010,mannella2012} helped.

Three practical rules follow. Infer the
diffusion tensor together with the drift, never as an afterthought:
a drift map paired with a constant $D$ is a wrong model. Measure
currents with mid-point averages, Eq.~\eqref{eq:mid}, and test for
nonequilibrium with time-antisymmetric statistics of the data, not
with the curl of a drift map or the dissipation of a fitted
model. Separate the spurious drift from the force with the
moments of Eq.~\eqref{eq:identity}, in or out of equilibrium, and
simulate with the explicit $\nabla\cdot\D$ rather than an end-point
scheme~\cite{depirey2022} (Appendix~B). Finally, I propose to reserve the name \emph{spurious drift} for the It\^o drift $\nabla\cdot\D$ that a thermal bath imposes when diffusion is inhomogeneous, as distinct from the noise-induced drifts that fast degrees of freedom or
self-generated viscosity fields can produce out of
equilibrium~\cite{volpe2016,pesce2013,cates2013,nakayama2026}: those
carry currents and do work, and the mid-point velocity tells them apart.
The spurious drift does neither. Consistently paired with the
state-dependent diffusion, it keeps the density Boltzmann-distributed
and the currents at zero. It can be measured, mapped, and mistaken for a
force, but it moves nothing. \emph{Eppur non si muove.}

\begin{acknowledgments}
  I thank Emmanuel Trizac, Martin Lenz, Howard
  Stone, and Matthieu Coppey for insightful discussion and comments,
  and Maxime Lavaud, Thomas Salez and Yacine Amarouchene for sharing
  their data. AI was used to finalize this article (Claude Opus 5.5;
  figures, code and text); all ideas and results presented here are
  mine. The project leading to this publication has received funding
  from France 2030, the French Government program managed by the
  French National Research Agency (ANR-16-CONV-0001) and from
  Excellence Initiative of Aix-Marseille University---A*MIDEX,
  ICTP-SAIFR (FAPESP Grant No.~2021/14335-0) where part of this work
  was done, and the European Union (ERC-SuperStoc-101117322).
\end{acknowledgments}

\bibliography{spurious_drifts}

\clearpage

\section*{Appendices}%
\label{app:appendices}

\subsection*{Appendix A: Discrete models.}
\label{app:chain}
\begin{figure}[b]
  \includegraphics[width=\columnwidth]{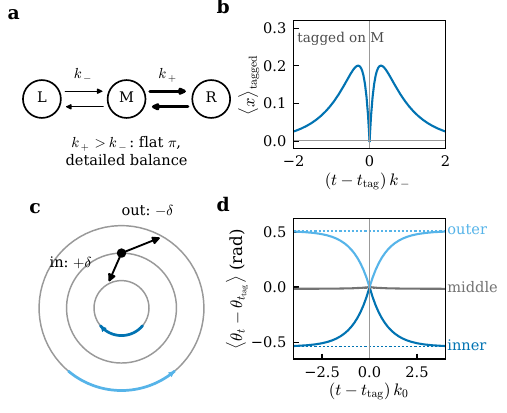}
  \caption{\label{fig:discrete}%
    \textbf{Discrete models.} (a)~Three-state chain, $k_+=3k_-$, and
    (b)~the exact mean position of the particles tagged on $M$.
    (c)~Three rings (radii $0.5$, $1$, $1.5$; $\delta=\pi/6$; rates
    $k_0\sqrt{\pi_j/\pi_i}$ in a harmonic trap); arcs: mean angular
    drift of the inner and outer rings. (d)~Exact mean rotation of a
    shell tagged on each ring; dashed: $-\delta(\av{i}-i_0)$.}
\end{figure}

\emph{Channel.} Three states $L, M, R$ at $x = -1, 0, +1$ exchange
particles at symmetric rates, $k_-$ on the slow bond and $k_+ > k_-$ on
the fast one (Fig.~\ref{fig:discrete}a). The flat distribution satisfies
detailed balance, yet a particle leaving $M$ goes right with
probability $k_+/(k_+ + k_-)$, and
\begin{equation}
  \label{eq:chain}
  \av{x_t | M} = (k_+ - k_-)\, e^{-(k_+ + k_-) t}\,
  \frac{\sinh(\Omega t)}{\Omega} ,
\end{equation}
with $\Omega = (k_+^2 - k_+ k_- + k_-^2)^{1/2}$, rises, peaks and
relaxes back to zero, identically forward and backward in time
(Fig.~\ref{fig:discrete}b): the skeleton of the ink drop and of
Fig.~\ref{fig:channel}c.

\emph{Exits.} In the continuous channel, the backward equation
$(D\psi')'=0$ gives the probability that a particle released at the
center first exits through the fast end,
\begin{equation}
  \label{eq:splitting}
  p_\rightarrow
  = \frac{\ln\left[1/(1-\gamma)\right]}{\ln\left[(1+\gamma)/(1-\gamma)\right]}
  = \frac{1}{2} + \frac{\gamma}{4} + O(\gamma^3) ,
\end{equation}
$0.82$ for $\gamma=0.95$. The current of
Eq.~\eqref{eq:langevin_equilibrium} is nonetheless $J = D'\rho -
(D\rho)' = -D\rho'$, Fick's law with a local
diffusivity~\cite{vanmilligen2005}, which vanishes for a flat profile:
by time reversal, entrances are as asymmetric as exits.

\emph{Rings.} A particle hops between three rings of radii $r_i$
(Fig.~\ref{fig:discrete}c), with detailed balance for the Boltzmann
weights $\pi_i$. Each outward hop rotates it by $-\delta$ and each
inward hop by $+\delta$, as steps along a tilted fast axis would:
$\theta + \delta\, i$ is conserved, the gearing of the continuous model.
The mean angular drift is
clockwise on the inner ring, counterclockwise on the outer one, and
averages to zero over $\pi$; a shell tagged on ring $i_0$ rotates by
$-\delta(\av{i_t|i_0} - i_0)$, even in time (Fig.~\ref{fig:discrete}d).

\subsection*{Appendix B: Three drifts, time reversal, and conventions.}
\label{app:threedrifts}
\emph{Three drifts.} In Eqs.~\eqref{eq:drifts}, the post-point drift is
Nelson's backward velocity~\cite{nelson1967,haussmann1986}, and the
mid-point velocity is the average of the pre- and post-point drifts. The
drift thus splits into a time-odd part, $\mathbf J/P$, and a time-even
part, $\nabla\cdot(\D P)/P$, the only one left at equilibrium.
Equations~\eqref{eq:drifts} hold as $\Delta t\to0$, but the
time-reversal argument holds at any $\Delta t$: at equilibrium the path
measure is invariant under reversal, so that $\hat v_{\rm post} = -\hat
v_{\rm pre}$ and every time-antisymmetric observable ($\hat v_{\rm
mid}$, winding, entropy production) averages to zero.

\emph{Identifiability.} Drift statistics determine only $\Phi$ and
$\nabla\cdot(\D P)/P$. They cannot fix $\nabla\cdot\D$, since any $\D
+ \mathbf{A}$ with $\nabla\cdot(\mathbf{A}P) = 0$ leaves them
unchanged: the spurious drift is accessible only through the quadratic
variation, $\av{\Delta\x\Delta\x^{\sf T}|\x} \to 2\D\Delta t$.

\emph{Conventions.} Evaluating the noise amplitude $\mathbf g$ ($\mathbf
g\mathbf g^{\sf T} = 2\D$) at the end point of each step adds $[\mathbf
g\nabla\mathbf g]_\nu = g_{\mu k}\partial_\mu g_{\nu k}$ to the
drift~\cite{lau2007}, which depends on the choice of $\mathbf g$. In 2D,
$\mathbf g = \mathbf S\mathbf R(\chi)$, with $\mathbf S$ the symmetric
root and $\mathbf R(\chi)$ an arbitrary rotation field, and $\mathbf
g\nabla\mathbf g = \mathbf S\nabla\mathbf S + \det\mathbf
S\,\boldsymbol\epsilon\nabla\chi$, where $\boldsymbol\epsilon = \mathbf
R(\pi/2)$ is the rotation by a quarter turn. For the spiral, the
symmetric root misses a radial drift $\boldsymbol\rho \equiv
\nabla\cdot\D - \mathbf S\nabla\mathbf S =
2D_0(1-\sqrt{1-a^2})\,\hat{e}_r/r$ outside the core, $63\%$ of the
vortex at $a=0.9$. A rotation field can restore it only if
$\nabla\cdot(\boldsymbol\rho/\det \mathbf S) = 0$, which fails for both melted-core defects (Sec.~S8). End-point schemes are thus unsafe for
tensor diffusivity~\cite{depirey2022}; the explicit $\nabla\cdot\D$, or
a kinetic stochastic integral~\cite{hutter1998}, should be used.

\subsection*{Appendix C: Defects, gearing and zero torque.}
\label{app:vortex}
\emph{Spiral.} With $Q_c = Q_{xx} + iQ_{xy} = \tfrac{s}{2}e^{2i\phi}$
and $\phi = k\theta+\phi_0$ the director angle,
$[\nabla\cdot\mathbf Q]_x + i[\nabla\cdot\mathbf Q]_y = (\partial_x -
i\partial_y) Q_c$, which gives Eq.~\eqref{eq:divQ}. For the spiral of
Fig.~\ref{fig:vortex} ($\phi_0=-\pi/4$, fast axis $45^\circ$ clockwise
from the radius), $D_{rr} = D_{\theta\theta} = D_0$, $D_{r\theta} =
-aD_0 s$, with $s = r^2/(r^2+\xi^2)$, and the It\^o drift is
\begin{equation}
  \label{eq:drift2D}
  \Phi = -D_0\, r\,\hat{e}_r
  + aD_0\, r\Big[s - \frac{2(r^2+2\xi^2)}{(r^2+\xi^2)^2}\Big]\hat{e}_\theta
\end{equation}
(units $\ell=1$): clockwise inside $r\simeq1.4\,\ell$, counterclockwise
outside. Read as a velocity it gives a mean angular speed
$\av{\Phi_\theta/r}_P \simeq -2.7$~rad per $\ell^2/D_0$, against a
measured mean winding of $-1.3\pm2.0$~rad after $200\,\ell^2/D_0$ (Sec.~S3).

\emph{Gearing.} For polar components of $\D$ depending on $r$ only and a
radial $V$, It\^o's lemma gives the angular drift $\omega =
(D_{r\theta}' - D_{r\theta}V')/r$; reading the drift as a velocity adds
a lever-arm term, $\Phi_\theta/r = \omega + 2D_{r\theta}/r^2$. The
variable $\theta - G(r)$, with $G' = D_{r\theta}/(rD_{rr})$, is a
martingale uncorrelated with $r$: the angle is geared to the radius, and
$\av{\theta_t - \theta_0|r_0} = \av{G(r_t)|r_0} - G(r_0)$, even in time.
For the spiral, $G = -\tfrac{a}{2}\ln(r^2+\xi^2)$ and $r$ is the radius
of an isotropic Ornstein--Uhlenbeck process, whose propagator gives the
black lines of Fig.~\ref{fig:vortex}d; tagged shells reverse their sense
at $r^*\simeq1.10\,\ell$. Those of Fig.~\ref{fig:vortex}j solve
$\partial_t P = \nabla\cdot[\D e^{-V}\nabla(P e^{V})]$ from the tagged disc (Sec.~S4).

\emph{Zero torque.} Writing $\Phi P = \mathbf J + \nabla\cdot(\D P)$ and
integrating by parts, the symmetry of $\D$ gives $\av{\x\times\Phi}_P =
\int\x\times\mathbf J\,\mathrm{d}\x$ in the absence of boundaries: the
torque of the drift is twice the area-enclosing rate of the
trajectories~\cite{ghanta2017,frishman2020}, and vanishes at
equilibrium. The spiral drift must therefore counter-rotate.

\emph{Active analogy.} The naive model $(\Phi,D_0\openone)$ is a tracer
in the potential $V$ with the extra drift $\beta
e^{V}\nabla\cdot(e^{-V}\mathbf Q)$. To linear order in $a$, the density
adjusts to screen the gradient part of this forcing, and the current is
the rest: $P_{\rm eq}\beta e^{V}\nabla\cdot(e^{-V}\mathbf Q) = \mathbf J
+ P_{\rm eq}\nabla\chi$ with $\nabla\cdot\mathbf J = 0$. This is the
single-particle analogue of the curvature current of dry active
nematics~\cite{ramaswamy2003}, with $\zeta=-\beta$ in the convention
where the active force is $-\zeta\nabla\cdot\mathbf Q$ ($\zeta>0$:
extensile). Far from a $+1/2$ core it is the jet $\mathbf J/P =
aD_0\cos\theta\,\hat e_r/r$, which the trap closes into two vortices
(Fig.~\ref{fig:vortex}k).

\subsection*{Appendix D: Pair hydrodynamics.}
\label{app:pair}
A direct computation gives Eq.~\eqref{eq:pairforce}. For the Oseen
relative tensor, $D_\parallel = 2D_0 - 4C/r$ and $D_\perp = 2D_0 -
2C/r$ ($C = k_{\rm B}T/8\pi\eta$): the spurious drift vanishes in
$d=3$, as for the Rotne--Prager correction, but not for motion
restricted to a plane, where $[\nabla\cdot\D_{\rm rel}]_r = 2C/r^2$ and
the naive potential is $\simeq(3R/4r)\,k_{\rm B}T$ for spheres of
radius $R$~\cite{bleibel2014}.

\emph{Depletion monolayer.} The data of Ref.~\cite{king2025} are $61$
videos of $139$ frames, $7.16$~ms apart. Tracked positions carry a
pairwise localization bias of $2$--$5$~nm, which is calibrated on rigid
dimers and removed (Sec.~S9). The moment-route well,
$-2.46\pm0.06\,k_{\rm B}T$ at $1.33\,\mu$m, matches $-\ln g$ ($-2.42$).
On uncorrected tracks it is $-2.60\pm0.06$, $-2.60\pm0.06$ and
$-2.60\pm0.16$ at $\Delta t = 7$, $57$ and $459$~ms, against $-0.61$,
$-0.29$ and $-0.01$ for the naive well. In Fig.~\ref{fig:pair}b,
$\hat\Phi_r$, $D_\parallel$ and $D_\perp$ are binned relative increments
of all pairs, and $\hat\sigma = \hat\Phi_r - \langle D_\parallel\hat
F\rangle$. Read with $2\Dbar$, the force part alone gives a well of
$-0.96\,k_{\rm B}T$; adding Eq.~\eqref{eq:pairforce} gives $-0.57$, the
naive value. Bound pairs relax by a quarter within a frame, which smears
the split inside the $40$-nm well: on simulated pairs with a known
lubrication profile, the integral of $\hat\sigma$ over the well is
recovered to $12\%$, but not its shape (Sec.~S9). Here it is
$0.035\pm0.001\,\mu$m$^2$/s, against $0.045$ from
Eq.~\eqref{eq:pairforce}.

\subsection*{Appendix E: Numerics and models.}
\label{app:numerics}
\emph{Simulations.} Euler--Maruyama with reflecting walls (1D)
and the SFI integrator~\cite{frishman2020,sfi_zenodo} (defects), all
with the explicit It\^o drift. Each defect ensemble comprises $100$
trajectories of $200\,\ell^2/D_0$ recorded every $10^{-3}\,\ell^2/D_0$;
their drift maps agree between movies and with the exact drift, their
mid-point maps with zero (Sec.~S1).

\emph{Models.} $\hat\Phi$ is the linear SFI estimator of the drift,
$\Dbar$ the average quadratic variation and $\Dhat(\x)$ its linear
fit on a tensor dictionary. The consistent model's drift,
$\nabla\cdot\Dhat + \Dhat\hat{\mathbf F}$, equals $\hat\Phi +
(\nabla\cdot\Dhat - \hat\sigma)$ up to projection error: the measured drift, corrected by the
mismatch between the differentiated fit and the moment-route spurious
drift ($5$--$7\%$ here). The control model omits this correction, and
its residual inconsistency drives a current: it winds at
$-0.08\pm0.03$~rad per $\ell^2/D_0$ around the spiral, and erratically
with noisy data (Sec.~S6). Naive models are evaluated by their
stationary Fokker--Planck solution (spiral: $-2.8$~rad per
$\ell^2/D_0$, $\dot S \simeq 3.7\,k_{\rm B}D_0/\ell^2$; $+1/2$: $\dot
S\simeq0.65$), the others by simulation (Sec.~S3).

\emph{Realistic regime.} With $a=0.25$ (diffusivity ratio $5/3$), $40$
trajectories sampled every $0.05\,\ell^2/D_0$ and localization noise
$0.2\,\ell$, the moment route still detects the spurious drift ($\nu =
0.16$, floor $0.04$); naive spiral models wind at $-0.4\pm0.2$, consistent
ones at $0.02\pm0.03$ (Sec.~S6).

\subsection*{Appendix F: The moment route.}
\label{app:moments}
Project Eq.~\eqref{eq:identity} on basis functions $\mathbf b_a(\x)$,
and expand the score on a second basis, $\nabla\ln P\simeq\sum_\alpha
c_\alpha\mathbf s_\alpha$. Each ingredient is an average over the
recorded steps:
\begin{align}
  \label{eq:moments}
  J_a &= \av{\bar{\mathbf b}_a\cdot\Delta\x}/\Delta t, &
  M_{ab} &= \av{\mathbf b_a\cdot\D_t\mathbf b_b}, \nonumber\\
  M'_{a\alpha} &= \av{\mathbf b_a\cdot\D_t\mathbf s_\alpha}, &
  c &= -(G^{\mathcal S})^{+}\av{\nabla\cdot\mathbf s},
\end{align}
with $\bar{\mathbf b}_a$ the average of $\mathbf b_a$ over the two ends
of the step, $\D_t = \Delta\x\Delta\x^{\sf T}/2\Delta t$ (or a
noise-robust stencil~\cite{vestergaard2014,frishman2020}), and
$G^{\mathcal S}_{\alpha\beta}=\av{\mathbf s_\alpha\cdot\mathbf
s_\beta}$ (score matching~\cite{hyvarinen2005}). The force is then
$\hat{\mathbf F} = \sum_a f_a\mathbf b_a$ with $f = M^{-1}(J + M'c)$, and
the spurious drift is the rest of the measured drift, $\hat\sigma =
\hat\Phi - \D\hat{\mathbf F}$, projected on the same basis. No
diffusion field is fitted or differentiated, $\hat{\mathbf F}$ is not
constrained to be a gradient, and the only approximations are the finite
bases; at equilibrium, $J=0$ and the method reduces to Boltzmann
inversion. The statistic $\nu = \|\hat\sigma\|/\|\hat\Phi\|$ is compared
with its value on a simulation of $(\hat\Phi,\Dbar)$ of the same
size (Sec.~S2). For the spiral, the $+1/2$ and the colloid above the wall, $\nu = 0.92$, $0.57$ and
$0.74$ (floors $0.018$, $0.022$ and $0.08$), and $\hat\sigma$ recovers
$\nabla\cdot\D$ to $7\%$, $5\%$ and $19\%$ (relative $L^2$ error; Secs.~S3 and~S10).

\clearpage

\setcounter{equation}{0}\setcounter{figure}{0}\setcounter{table}{0}
\renewcommand{\theequation}{S\arabic{equation}}
\renewcommand{\thefigure}{S\arabic{figure}}
\renewcommand{\thetable}{S\arabic{table}}
\renewcommand{\theHequation}{S\arabic{equation}}
\renewcommand{\theHfigure}{S\arabic{figure}}
\renewcommand{\theHtable}{S\arabic{table}}
\section*{Supplemental Material}
This Supplemental Material gives the simulation and inference
settings, the equilibrium checks and the complete tables behind
Figs.~2 and~3 and the numbers quoted in the main text and its appendices.
Equation, figure and appendix labels without the prefix S refer to the main text; $k_{\rm
B}T=1$. For the defects, lengths are in units of the trap length $\ell$
and times in $\ell^2/D_0$, so that winding rates are in rad per
$\ell^2/D_0$, the lobe circulation $A$ in $D_0$ and the entropy
production $\dot S$ in $k_{\rm B}D_0/\ell^2$. A number in parentheses is
one standard error on the last digit. Every number is regenerated by the
reproducibility package (Sec.~S11).

\section*{S1. Equilibrium ensembles around the defects}
Each defect ensemble of Fig.~2, from which the drift maps are binned
and all models inferred, comprises $100$ trajectories of
$200\,\ell^2/D_0$ recorded every $10^{-3}\,\ell^2/D_0$ (SFI
integrator~\cite{sfi_zenodo}, explicit It\^o drift of Eq.~(1)); the ``reversed
movie'' is the same ensemble read backwards. Table~\ref{tab:chi2}
checks that they are at equilibrium, so that the winding of the
naive models (Fig.~2f,l) is attributable to the inference alone (Appendix~E):
drift maps binned in squares of side $0.5\,\ell$ within $r<2.4\,\ell$
($152$ degrees of freedom) are compared with the exact drift of
Eq.~(1), with each other and, for the mid-point velocity, with zero.
\begin{table}[h]
\caption{\label{tab:chi2}Equilibrium checks on the binned drift
maps, $\chi^2$ per degree of freedom.}
\begin{tabular}{lcc}
\hline
 & spiral & $+1/2$\\
\hline
forward drift vs exact $\Phi$ & 1.05 & 1.00\\
reversed drift vs exact $\Phi$ & 1.32 & 0.93\\
forward vs reversed drift & 0.90 & 0.93\\
mid-point velocity vs 0 & 0.74 & 1.08\\
\hline
\end{tabular}
\end{table}
The binned stationary density deviates from $e^{-V}$ by a median
relative $3.4\%$ (spiral) and $6.2\%$ ($+1/2$); the winding and lobe
circulation of the data are compatible with zero (Table~\ref{tab:models}).

\section*{S2. Inference settings}
The estimators of Appendix~E and the moment route of Appendix~F were run on the
defect ensembles with the following bases, time steps and decision
rules, on which every inferred model of Fig.~2 and every value of
$\nu$ depend.

\emph{Dictionaries.} Scalar radial functions are Chebyshev polynomials
$T_n(u)$ of $u = \ln\rho/1.8$, $\rho = (r^2+0.2^2)^{1/2}$.
\emph{Spiral} (rotation-covariant): vectors
$T_n\,\x/\rho$ and $T_n\,\hat e_z\times\x/\rho$; tensors
$T_n\openone$ and $T_n$ times the two winding-2 traceless tensors
built from $(x^2-y^2, 2xy)/\rho^2$. Degrees: $5$ for the drift and
force, $10$ for the score, $5$ for $\D$. \emph{$+1/2$} (no symmetry
assumed): $T_n$ times the Cartesian unit vectors times $\{1, x/\rho,
y/\rho, (x^2-y^2)/\rho^2, 2xy/\rho^2\}$; degrees $4$ (drift, force),
$6$ (score), $2$ ($\D$, three tensor components). \emph{Realistic
spiral} (Sec.~S6): degrees $3$, $6$, $3$; degree $5$ gave an
ill-conditioned fit on the noisy data.

\emph{Naive and control models.} $\hat\Phi$ is the linear SFI
estimator~\cite{frishman2020} on the drift dictionary; $\Dbar$ the average quadratic
variation; $\Dhat(\x)$ the linear SFI fit of the quadratic variation on
the tensor dictionary. All fits use the first $50\,\ell^2/D_0$ of each
of the $100$ trajectories.

\emph{Moment route.} The same data are subsampled to $\Delta t = 5\times
10^{-3}\,\ell^2/D_0$. On clean data, $\D_t$ is the forward weak-noise
stencil of Ref.~\cite{frishman2020}. On the realistic data of Sec.~S6, where the code
detects localization noise (bias-corrected level $0.95\,{\rm
Tr}\Dbar\Delta t$), $\hat\sigma$ and $\nu$ use the drift-corrected raw
stencil and $\hat{\mathbf F}$ the denoised pair. The norms entering
$\nu$ are restricted to the data points where $|\nabla\hat\Phi|\Delta t$
is at most twice its root mean square ($95\%$ of the data for the
spiral, $93.9\%$ for the $+1/2$, all of the realistic data). Standard
errors come from blocks of trajectories. The floor quoted with each
$\nu$ is its value on one simulation of $(\hat\Phi,\Dbar)$ with the size
and sampling of the data ($10$ integrator substeps per frame), i.e.\ the
level produced by fitting error alone. Detection requires $\nu -
\nu_{\rm floor}$ to exceed both $3\sqrt2\,\delta\nu$ and $0.25\,\nu_{\rm
floor}$.

\emph{Inference variance.} Each of $20$ bootstrap replicates resamples
the analyzed trajectories with replacement and reruns the whole
pipeline on the forward movie (naive, control and moment-route fits,
the latter without null floor), then simulates the three models. The
standard deviation over replicates (``bootstrap s.d.'', the error bar
on inferred models in Fig.~2f,l) includes the simulation noise of one
model run.

\section*{S3. Model comparisons}
Table~\ref{tab:models} completes Fig.~2f,l (one number per model)
with the reversed movie, the entropy production and both error
estimates; Table~\ref{tab:mr} holds the moment-route diagnostics
quoted in Appendix~F. Models are simulated from $100$ Boltzmann-distributed
initial conditions for $200\,\ell^2/D_0$ (frames every
$10^{-2}\,\ell^2/D_0$), with standard errors over these trajectories
at fixed inferred fields. The naive models are also evaluated exactly,
by their stationary Fokker--Planck (FP) solution in the reflecting
box $|x|,|y|\le4\,\ell$ (finite volumes, $h=0.05\,\ell$), the values
shown in Fig.~2f,l for the reasons given below.
\begin{table*}[tb]
\caption{\label{tab:models}Time-antisymmetric observables of the data
and of the inferred models (Fig.~2f,l), forward / reversed movie.
Upper block: standard errors over $100$ simulated trajectories at
fixed fields. Lower block: bootstrap s.d.\ over $20$ replicates of the
whole pipeline (forward movie).}
\begin{tabular}{lcccc}
\hline
 & \multicolumn{2}{c}{spiral} & \multicolumn{2}{c}{$+1/2$}\\
model & winding & $\dot S$ & $A$ & $\dot S$\\
\hline
naive, FP & $-2.75$ / $-2.80$ & $3.73$ / $3.82$ & $0.333$ / $0.331$ & $0.657$ / $0.637$\\
naive, simulated & $-2.47(3)$ / $-2.48(3)$ & $3.33$ / $3.36$ & $0.25(1)$ / $0.30(1)$ & $0.61$ / $0.60$\\
naive, exact $\Phi$ & $-2.711$ & $3.668$ & $0.32$--$0.33$ & $0.60(1)$\\
control $(\hat\Phi,\Dhat)$ & $-0.08(2)$ / $-0.11(2)$ & $0.009$ / $0.016$ & $0.003(3)$ / $0.006(3)$ & $0.004$ / $0.008$\\
consistent (moment route) & $0.005(16)$ / $-0.026(16)$ & $\lesssim0.003$ & $0.002(2)$ / $0.002(3)$ & $0.003(5)$ / $0.008(5)$\\
true model & $-0.009(16)$ & $0.002$ & $0.001(3)$ & $0.001(5)$\\
data & $-1.3\pm2.0$\,rad at $t=200$ & & $0.003(3)$ & \\
\hline
\multicolumn{5}{l}{bootstrap s.d.}\\
naive, simulated & $0.046$ & $0.075$ & $0.116$ & $0.085$\\
naive, FP & $0.052$ & $0.066$ & $0.014$ & $0.034$\\
control & $0.028$ & $0.006$ & $0.003$ & $0.002$\\
consistent & $0.016$ & $0.002$ & $0.005$ & $0.002$\\
\hline
\end{tabular}
\end{table*}

\emph{Spiral.} The naive model built on the exact drift winds at
$\av{\Phi_\theta/r}_P = -2.711$, $74.5\%$ of it from $r<2\xi$; the
fitted ones at $-2.75$ and $-2.80$ (FP; bootstrap $-2.77\pm0.05$).
Their simulated windings are lower, $-2.47$ and $-2.48$, because
angles recorded every $10^{-2}\,\ell^2/D_0$ alias near the core,
where the angular velocity is largest: the exact naive model simulated
with the same protocol winds at $-2.44\pm0.06$, and at $-2.63\pm0.07$
with frames every $10^{-3}$. This bias is common to all simulated
models and does not affect those that do not wind. The control model
winds $5$--$7$ standard errors from zero at fixed fields and $2.5$
bootstrap s.d.\ (bootstrap mean $-0.069$): its $\hat\Phi$ and $\Dhat$
are fitted independently, and their residual inconsistency drives a
current (Appendix~E). The concentration of the naive winding in the core
is generic: the mean angular drift $\av{\omega}_P = -2\pi
D_{r\theta}(0)P(0)$ is a boundary term that the melted core cancels,
so that $\av{\Phi_\theta/r}_P$ comes entirely from the lever-arm term
$2D_{r\theta}/r^2$ of Appendix~C and diverges logarithmically as $\xi\to0$.

\emph{$+1/2$.} The FP stationary state of the fitted naive models gives
$A = 0.333$ and $0.331$, $\dot S = 0.657$ and $0.637$, close to the
exact-drift values. The simulated values are lower and differ between
movies beyond their standard errors because the unconfined simulated
trajectories reach regions beyond the data, where the fitted drift is
extrapolated (mean position $-1.33$ and $-0.94\,\ell$, against the FP
value $-0.33\,\ell$). Over bootstrap replicates the simulated naive
$A$ scatters widely ($0.23\pm0.12$; on average $18\%$ of the walkers
escape) whereas the FP value is stable, $A = 0.327\pm0.014$, $\dot S =
0.66\pm0.03$; Fig.~2l therefore shows the latter. The control and
consistent models circulate at $0.001\pm0.003$ and $-0.002\pm0.005$
(bootstrap mean $\pm$ s.d.).

\emph{Moment route.} Detected from both movies (Table~\ref{tab:mr}); $\hat\sigma$
recovers $\nabla\cdot\D$ to $7\%$ (spiral) and $5\%$ ($+1/2$, lower
half of Fig.~2g), $\hat{\mathbf F}$ recovers $-\x$ to $8$--$10\%$ and
$15$--$19\%$, the naive force being off by more than $100\%$.
\begin{table}[h]
\caption{\label{tab:mr}Moment-route diagnostics, forward / reversed
movie: $\nu$ (block standard error) and its null floor; relative
$L^2$ errors over the data points within $r<2.5\,\ell$ of
$\hat\sigma$, of $\hat{\mathbf F}$ and of the naive force.}
\resizebox{\columnwidth}{!}{%
\begin{tabular}{lccc}
\hline
 & spiral & $+1/2$ & realistic (S6)\\
\hline
$\nu$ & $0.919(7)$ / $0.913(4)$ & $0.572(4)$ / $0.572(3)$ & $0.161(3)$ / $0.160(2)$\\
floor & $0.018$ / $0.015$ & $0.022$ / $0.019$ & $0.037$ / $0.037$\\
$\hat\sigma$ & $7.2\%$ / $7.3\%$ & $5.2\%$ / $5.2\%$ & $47\%$ / $48\%$\\
$\hat{\mathbf F}$ & $7.6\%$ / $10\%$ & $19\%$ / $15\%$ & $4.6\%$ / $3.8\%$\\
naive $\Dbar^{-1}\hat\Phi$ & $144\%$ / $146\%$ & $107\%$ / $107\%$ & $93\%$ / $92\%$\\
\hline
\end{tabular}}
\end{table}

\section*{S4. Tagged ensembles}
The black curves of Fig.~2d,j are the parameter-free predictions of
Eq.~(1) for the conditioned rotation and push described in Appendix~C,
forward and backward in time. \emph{Spiral (Fig.~2d):} the gearing
relation of Appendix~C, $\av{\theta_t-\theta_0\,|\,r_0} = \av{G(r_t)\,|\,r_0}
- G(r_0)$ with $G = -\tfrac a2\ln(r^2+\xi^2)$, is evaluated by
quadrature with the propagator of the isotropic Ornstein--Uhlenbeck
process of which $r$ is the radius; the plateau,
$\av{G}_{\rm Boltzmann} - G(r_0)$, changes sign at $r^* \simeq
1.10\,\ell$ and matches the data within $1.2$~SEM for $r_0 =
0.3$--$2\,\ell$. \emph{$+1/2$ (Fig.~2j):} the Fokker--Planck equation
in equilibrium form, $\partial_t P = \nabla\cdot[\D
e^{-V}\nabla(Pe^{V})]$, solved on a grid ($h = 0.05\,\ell$) from the
Boltzmann density restricted to the tagged disc $r<0.15\,\ell$; the
data lie within $0.75$~SEM (root mean square). The spurious drift of
Eq.~(4) pushes through the core at up to $1.1\,aD_0/\xi \simeq
4\,D_0/\ell$, and
the tagged particles move by up to $0.6\,\ell$ towards the tail
before relaxing back.

\section*{S5. Out of equilibrium: a hidden rotation}
The nonequilibrium spiral of the main text (data winding at $2.7$~rad
per $\ell^2/D_0$, naive model at $0.3$, consistent at $2.6$) is the
ideal spiral ($a=0.9$) advected by the flow
$\mathbf u = \Omega\hat e_z\times\x$, i.e.\ subjected to the drag force
$\D^{-1}\mathbf u$, so that $\Phi = \nabla\cdot\D - \D\nabla V +
\mathbf u$. Since $\mathbf u$ is azimuthal and the spiral is
rotation-covariant, $\nabla\cdot(e^{-V}\mathbf u) = 0$: the stationary
density remains $e^{-V}$ (measured $\av{r^2} = 1.99$) and the current
$e^{-V}\mathbf u$ rotates rigidly. With $\Omega = 2.711 =
-\av{\Phi^{\rm eq}_\theta/r}_P$, the density-weighted mean angular
drift vanishes: read as a velocity, the drift map shows no rotation
(measured $\av{\hat\Phi_\theta/r}_P = 0.08$). Same protocol as
Sec.~S1 ($100$ trajectories, $\Delta t = 10^{-3}$, first
$50\,\ell^2/D_0$ analyzed); simulated windings share the sampling bias
of Sec.~S3, which does not affect the comparison (Table~\ref{tab:torque}).
\begin{table}[h]
\caption{\label{tab:torque}Hidden rotation: winding rates (fields from
the forward / reversed movie) and entropy production (forward).}
\begin{tabular}{lcc}
\hline
model & winding (fwd / rev) & $\dot S$ (fwd)\\
\hline
data & $+2.72(3)$ & $14.5$\\
true & $+2.69(2)$ & $14.4$\\
naive $(\hat\Phi,\Dbar)$ & $+0.32(3)$ / $-5.16(3)$ & $17.5$\\
Boltzmann $(\mathbf F = \nabla\ln P)$ & $-0.02(2)$ / $0.00(2)$ & $0.002$\\
control $(\hat\Phi,\Dhat)$ & $+2.68(1)$ / $-2.81(2)$ & $14.4$\\
consistent (moment route) & $+2.58(2)$ / $-2.48(2)$ & $10.4$\\
\hline
\end{tabular}
\end{table}
The moment route detects the spurious drift ($\nu = 0.46$, floor
$0.012$; reversed: $0.44$, $0.015$), recovers $\nabla\cdot\D$ to $8\%$
and the force, trap plus drag, to $6\%$ ($7\%$ from the reversed
movie, whose force is $-\x - \D^{-1}\mathbf u$). $\hat{\mathbf F}$ is
biased low, mostly near the core: by $27$, $13$ and $6\%$ (radial)
and $13$, $7.5$ and $2\%$ (azimuthal) at $r = 0.5$, $1$ and $2\,\ell$,
where the drag is $3.5$, $12.6$ and $34\,k_{\rm B}T/\ell$ (angle
averages on circles). This lowers the winding of the consistent model
by $4\%$ and its entropy production, quadratic in the force, by
$28\%$. Boltzmann inversion ignores the current and returns the
equilibrium model; the naive model realizes the drift-as-velocity
reading; the control model, a correct generative model when $\Dhat$
is accurate, reproduces the dynamics but does not separate the force
from the spurious drift.

\section*{S6. Realistic regime}
The robustness claim of the main text and the numbers of Appendix~E ($\nu =
0.16$, windings $-0.4\pm0.2$ and $0.02\pm0.03$) rest on the spiral
with $a = 0.25$ ($D_\parallel/D_\perp = 5/3$), $100$ trajectories
simulated for $500\,\ell^2/D_0$ of which $40$ are analyzed, sampled
every $0.05\,\ell^2/D_0$ with Gaussian localization noise of standard
deviation $0.2\,\ell$ per coordinate. The moment route detects the
spurious drift from both movies ($\nu = 0.16$, floor $0.037$;
Table~\ref{tab:mr}); at this noise level $\hat\sigma$ is recovered
only to $47\%$, but the force, which dominates the drift, to
$4$--$5\%$ (naive: $93\%$). The naive models wind at $-0.31(2)$ and
$-0.54(2)$ (forward / reversed), with $\dot S = 0.10$ and $0.19$; the
consistent ones at $0.02(2)$ and $-0.01(2)$, with $\dot S = 0.0023$ and
$0.0032$. The naive rates differ between movies because the naive
winding is dominated by the core, which the noisy fits resolve
differently (fitted $\av{\Phi_\theta/r}_P = -0.34$ and $-0.65$,
against $-0.75$ exactly); over bootstrap replicates the naive rate is
$-0.41\pm0.18$, so the two movies agree within the inference variance,
and the consistent models wind at $0.015\pm0.029$ ($\dot S =
0.005\pm0.002$). The control model $(\hat\Phi,\Dhat)$, with parametric
noise-robust fits, is erratic: $0.18\pm0.22$, $\dot S = 0.05\pm0.05$.
Experimentally, a tracer larger than the core samples an averaged
$\D$ and its radius sets the effective $\xi$; the spurious drift,
$\sim aD_0/r$, matters where genuine forces are of order $k_{\rm B}T$
and is negligible against the elastic energies of anchored colloids,
$\sim KR \gtrsim 10^3\,k_{\rm B}T$ for a Frank constant $K$ and a
colloid radius $R$.

\section*{S7. Scalar-diffusivity inference}
The main text states that scalar-diffusivity inference would not help
around the defects, ${\rm Tr}\,\D$ being uniform; here this is
verified with TRamWAy~0.7~\cite{serov2020}, including its Bayes-factor test
designed to tell a genuine force from a spurious $\nabla D$. Settings:
regular grid ($\sim2000$ points per cell), maximum-a-posteriori
inference of a scalar diffusivity and a force per cell (``df''),
cells with more than $500$ points and $r<2.2\,\ell$. For Eq.~(3),
${\rm Tr}\,\D = 2D_0$, the scalar correction $\nabla D$ vanishes and
the whole It\^o drift is read as a force. \emph{Ideal spiral}
(subsampled to $\Delta t = 0.05\,\ell^2/D_0$): the radial force
matches the trap, $\av{F_r/(-r)} = 0.998$, but the azimuthal force is
the spurious vortex, $\av{F_\theta} = -2.38\pm0.05$, $-0.98\pm0.13$ and
$+0.52\pm0.08\,\ell^{-1}$ for $0.2\le r<0.8$, $0.8\le r<1.4$ and
$1.4\le r<2.2\,\ell$; root mean square $F_\theta/F_r = 0.69$. The Bayes
test finds a genuine force in all $40$ cells (median $\log_{10}B =
85$). \emph{Realistic spiral:} $\av{F_\theta} = -0.52$, $-0.27$,
$+0.17\,\ell^{-1}$ in the same shells, matching $\Phi_\theta/D_0$; the
radial force is overestimated by a factor $1.85$ (localization
noise); genuine force in all $32$ cells (median $\log_{10}B = 149$).

\section*{S8. End-point conventions}
The criterion of Appendix~B, by which no square root $\mathbf g$ of
$2\D$ makes the end-point convention reproduce $\nabla\cdot\D$ around
the defects, follows from the fact that evaluating $\mathbf g$ at the
end point adds $[\mathbf g\nabla\mathbf g]_\nu =
g_{\mu k}\partial_\mu g_{\nu k}$ to the It\^o drift~\cite{lau2007}. In 2D every
square root of $2\D$ is $\mathbf g = \mathbf S\mathbf R(\chi)$, with
$\mathbf S$ the symmetric root and $\mathbf R(\chi)$ a rotation by a
field $\chi(\x)$. Since $\partial_\mu\mathbf R\,\mathbf R^{\sf T} =
\boldsymbol\epsilon\,\partial_\mu\chi$, with $\boldsymbol\epsilon =
\mathbf R(\pi/2)$ the rotation by a quarter turn, and $\mathbf
S\boldsymbol\epsilon\mathbf S = \det\mathbf S\,\boldsymbol\epsilon$,
\begin{equation}
  [\mathbf g\nabla\mathbf g]_\nu = [\mathbf S\nabla\mathbf S]_\nu +
  \det\mathbf S\,[\boldsymbol\epsilon\nabla\chi]_\nu ,
\end{equation}
with $[\mathbf S\nabla\mathbf S]_\nu = S_{\mu\rho}\partial_\mu
S_{\nu\rho}$ (checked numerically to $2\times10^{-16}$). The end-point
drift equals $\nabla\cdot\D$ if and only if $\det\mathbf
S\,\boldsymbol\epsilon\nabla\chi = \boldsymbol\rho \equiv \nabla\cdot\D
- \mathbf S\nabla\mathbf S$, i.e.\ if
$\boldsymbol\epsilon^{\sf T}\boldsymbol\rho/\det\mathbf S$ is a
gradient, which requires $\nabla\cdot(\boldsymbol\rho/\det\mathbf S)
= 0$. This divergence does not vanish for the $+1$ spiral or the
$+1/2$ with $s = r^2/(r^2+\xi^2)$ (polar grid, $0.2\le r\le2\,\ell$):
no $2\times2$ square root works. For the spiral, $\boldsymbol\rho$ is
radial and reaches $63\%$ of the spurious vortex far from the core,
the figure quoted in Appendix~B. Only without a core ($\xi\to0$) does a convention exist:
$\boldsymbol\rho = 2D_0(1-\sqrt{1-a^2})\,\hat e_r/r$ is cancelled by
$\chi = -c\theta$, $c = (1-a^2)^{-1/2}-1$ ($1.294$ at $a=0.9$), a root
single-valued (up to the sign of $\mathbf g$) only if $2c$ is an
integer. Noise with more than two components (non-square
$\mathbf g$) is not covered.

\section*{S9. Colloid pairs in a depletion monolayer (Fig.~3a--c)}
Figure~3a--c and the well depths quoted in the main text and in Appendix~D come
from a reanalysis of the videos of Ref.~\cite{king2025}. This section gives the
data, the tracking and its bias correction, the inference settings, and
the calibration of the drift split at finite frame interval.

\emph{Data.} Ref.~\cite{king2025}, Harvard Dataverse doi:\allowbreak10.7910/\allowbreak DVN/\allowbreak HL2LD1
(CC0): $61$ bursts of $139$ frames, $7.164$~ms apart, recorded every
$30$~s; $1024^2$ pixels of $0.1767\,\mu$m ($181\,\mu$m field of view);
sulfate-polystyrene spheres of diameter $1.3\,\mu$m sedimented on glass
with a NaCMC depletant ($R_g\simeq60$~nm, size ratio $q\simeq0.09$, for
which the Asakura--Oosawa interaction is strictly pairwise); $2274$
particles per frame on average ($2032$--$2439$).

\emph{Tracking.} trackpy~0.7 on the raw frames
(\texttt{locate}: diameter $7$~px, \texttt{minmass} $5\times10^4$,
separation $5$~px; diameter $5$~px pixel-locks; \texttt{link}: search
range $2$~px, no memory); undetected entries are masked. The
localization noise of a pair separation is at most $2.4$~nm (lag-one
length increments of $843$ stuck dimers), negligible against
$\sqrt{2D_\parallel\Delta t}\simeq20$~nm. 

\emph{Localization bias.} A configuration-dependent bias is not
negligible: a neighbour's light inside the centroid mask shifts each
position by $\delta(r)$ along the line of centers. A consistency test
reveals it: the Boltzmann identity holds for test functions weighted by
any smooth function of the configuration, yet on raw tracks weighting by
coordination moves $U(1.6\,\mu{\rm m})$ from $-0.11$ to $-0.59$. The
bias is calibrated on $845$ rigid dimers (bonded in at least $60$
frames, length s.d.\ below $12$~nm), whose fixed length, regressed on
the positions of third particles, gives $\delta(r)$ on hat functions,
jointly with the weighting test (Table~\ref{tab:delta}).
\begin{table}[h]
\caption{\label{tab:delta}Pairwise localization bias $\delta(r)$;
negative values push pairs apart.}
\begin{tabular}{lcccccc}
\hline
$r$ ($\mu$m) & $1.27$ & $1.33$ & $1.39$ & $1.46$ & $1.55$ & $1.65$\\
$\delta$ (nm) & $-12.7$ & $-4.1$ & $-4.5$ & $-5.6$ & $-4.0$ & $-2.0$\\
\hline
$r$ ($\mu$m) & $1.77$ & $1.90$ & $2.05$ & $2.25$ & $2.45$ & \\
$\delta$ (nm) & $+1.9$ & $+3.5$ & $+1.5$ & $-3.3$ & $-1.2$ & \\
\hline
\end{tabular}
\end{table}
After the correction $x_i \to x_i - \sum_j\delta(r_{ij})\hat u_{ij}$, all
weightings agree to $0.02$, the shoulder disappears
($U(1.6\,\mu{\rm m})$: $-0.39\to-0.04$) and the well moves from $-2.60$
to $-2.46$ (Table~\ref{tab:king}); $-\ln g$ changes by $0.03$.

\emph{Inference.} SFI~\cite{sfi_zenodo}, multiplicative-noise module. Each burst is
a data set over all track identities with a validity mask and, as
extras, the list of pairs closer than $3\,\mu$m in any of its frames.
Basis: two uniform components and $18$ radial hat functions on pairs
(knots $1.12$, $1.18$, $1.22$, $1.26$, $1.29$, $1.32$, $1.35$, $1.38$,
$1.41$, $1.45$, $1.50$, $1.56$, $1.64$, $1.75$, $1.9$, $2.1$, $2.4$,
$2.7$ and $3.0\,\mu$m), multiplied by a window $w(\x_i)$ that vanishes
within $9\,\mu$m of the edge of the field and rises as $\sin^2$ over
$5\,\mu$m; its gradient enters $\av{\nabla\cdot\mathbf s}$, so no
boundary term arises. Weak-noise stencil, all particle pairs in $M$;
moments are summed over bursts and solved once; errors are jackknife
estimates over the $61$ bursts. For larger $\Delta t$, each burst is
split into $s$ interleaved sub-bursts ($s = 8$, $64$). $\Dbar =
0.0644\,\mu$m$^2$/s; potentials are $-\int\hat F_r\,\mathrm{d}r$, zero
at $3\,\mu$m. An independent implementation of the same estimator gives
the same wells to $0.05$.

\begin{table}[h]
\caption{\label{tab:king}Monolayer: well depths ($k_{\rm B}T$; jackknife
standard error on the last digit), at $1.33\,\mu$m (corrected tracks)
or $1.34\,\mu$m (raw tracks). $-\ln g$: $-2.42(1)$ corrected, $-2.45(2)$
raw.}
\begin{tabular}{lcccc}
\hline
tracks & $\Delta t$ (ms) & moment route & score & naive \\
\hline
corrected & $7.16$ & $-2.46(6)$ & $-2.49(5)$ & $-0.58(2)$\\
raw & $7.16$ & $-2.60(6)$ & $-2.62(5)$ & $-0.61(2)$\\
raw & $57.3$ & $-2.60(6)$ & $-2.63(5)$ & $-0.29(2)$\\
raw & $459$ & $-2.60(16)$ & $-2.63(5)$ & $-0.01(2)$\\
\hline
\end{tabular}
\end{table}

\emph{Drift split (Fig.~3b).} Relative increments of all pairs whose
members lie at least $3.2\,\mu$m from the edge are binned by the
separation at the start of the step ($20$-nm bins below $1.5\,\mu$m,
$40$-nm bins for $\hat\sigma$). $\hat\Phi_r = \av{\Delta\mathbf r\cdot\hat r}/\Delta t$;
$D_\parallel$ and $D_\perp$ are the conditional variances of the radial
and tangential increments divided by $2\Delta t$; the force part is
$D_\parallel$ times the moment-route force averaged over the pairs of
the bin; $\hat\sigma = \hat\Phi_r - \av{D_\parallel\hat F}$; and
Eq.~(6) is evaluated on the measured diffusivities. Over
$1.28<r<1.44\,\mu$m, $\int\hat\sigma\,\mathrm{d}r = 0.035(1)\,\mu$m$^2$/s
against $0.045(1)$ from Eq.~(6); isolated pairs (no third particle
within $3\,\mu$m) give $0.036(2)$ and $0.049(1)$. Integrated with
$2\Dbar$, the force part alone gives a well of $-0.96$, force part plus
Eq.~(6) $-0.57$, and the measured drift $-0.57$ (moment-route
estimator of the naive well: $-0.58$).

\begin{figure}[h]
\includegraphics[width=\columnwidth]{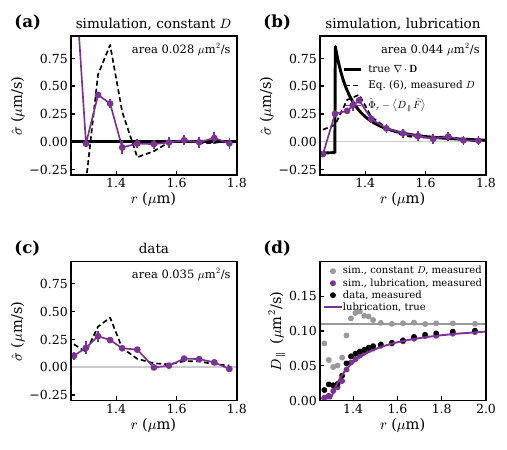}
\caption{\label{fig:kingcal}Calibration of the binned split at
$\Delta t = 7.164$~ms. (a,b)~Simulated isolated pairs with the
measured potential and constant $D$ (a) or a lubrication profile (b):
$\hat\sigma$ in $40$-nm bins ($\pm2$~SE), Eq.~(6) on the measured
diffusivities (dashed) and the exact $\nabla\cdot\D$ (solid).
(c)~Data, corrected tracks. (d)~Measured $D_\parallel$ in the three
cases and the imposed lubrication profile. Areas: $\int\hat\sigma\,\mathrm{d}r$
over $1.28$--$1.44\,\mu$m.}
\end{figure}

\emph{Finite frame interval.} Bound pairs explore a well of
s.d.\ $30$~nm and relax by about a quarter per frame (the
autocovariance of $r$ falls from $926$ to $667$~nm$^2$ in one frame).
Pre-point drifts and conditional variances are therefore attenuated
inside the well: from $7$ to $29$~ms, $\hat\Phi_r$ at $1.37\,\mu$m falls
from $-1.08$ to $-0.41\,\mu$m/s and $D_\parallel$ at $1.35\,\mu$m from
$0.030$ to $0.018\,\mu$m$^2$/s. The moment-route force, obtained from
configurations, is not attenuated. The split is calibrated on
simulated isolated pairs (Fig.~\ref{fig:kingcal}): relative coordinate
in $d=2$ with the measured $U(r)$ and either a constant
$D = 0.11\,\mu$m$^2$/s or $D_\parallel = 0.115\,h/(h+0.113)$,
$D_\perp = 0.13\,\mu$m$^2$/s ($h = r - 1.30\,\mu$m $+\,3$~nm), It\^o
Euler--Maruyama with the exact $\nabla\cdot\D$ and $20\,\mu$s steps,
sampled every $7.164$~ms with $1.7$~nm noise, $61$ videos of $441$
pairs, analyzed by the same code. The lubrication case reproduces the
measured $D_\parallel$ of the data and the integral of $\hat\sigma$
($0.044$ against the exact $0.050\,\mu$m$^2$/s), but not its shape:
$\hat\sigma$ peaks inside the well, whereas $\nabla\cdot\D$ is largest at
contact, where $D_\parallel$ is underestimated by about $25\%$. With
constant $D$, confinement within a frame alone yields $\int\hat\sigma =
0.028$ and $0.055$ from Eq.~(6); this case is excluded by the measured
$D_\parallel$ in the well ($0.02$--$0.03$, against $0.05$--$0.10\,\mu$m$^2$/s
for constant $D$).

\emph{Comparison with Ref.~\cite{king2025}.} Ref.~\cite{king2025} infers the potential from
the drift with a constant friction coefficient chosen separately; the
depth it obtains scales with that choice. With the friction set by the
measured mean diffusivity, this estimator gives $-0.6$ here.

\section*{S10. Colloid above a wall (Fig.~3d--f)}
The trajectory of Ref.~\cite{lavaud2021} reanalyzed in Fig.~3d--f (repository of
M.~Lavaud, MIT licence) is a polystyrene sphere above a glass plate,
$60$ frames per second, $115\,000$ frames, positions from Lorenz--Mie
holographic fits with the radius $r_p = 1.516\,\mu$m, refractive index
$1.597$ and pixel size $51.3$~nm fixed to the values recorded with the
trajectory file (the published calibration quotes
$1.518\pm0.006\,\mu$m, $1.584\pm0.006$ and $51.6$~nm/pixel~\cite{lavaud2021}). The
reference model of Fig.~3e has no free parameter once the wall
position $z_w$, the bulk diffusivity $D_0 = 0.145\,\mu$m$^2$/s, the
gravitational length $\ell_B = 0.471\,\mu$m, the Debye length $\ell_D
= 29.5$~nm and the electrostatic amplitude $B = 19.2\,k_{\rm B}T$ are
recalibrated on the data (\texttt{realdata/lavaud/recalib.json}):
$F(h) = (B/\ell_D)e^{-h/\ell_D} - 1/\ell_B$ with $h = z - z_w$, and
$D_\perp(h) = D_0(6h^2 + 2r_ph)/(6h^2 + 9r_ph + 2r_p^2)$, Brenner's
approximation~\cite{brenner1961}. The localization noise, measured on a stuck
particle, is $8.6$~nm in $z$. The measured $D_\perp$ and $D_\parallel$
(Fig.~3d) are quadratic variations in $18$ bins of $h$, $D_\parallel$
compared with Fax\'en's approximation as in Ref.~\cite{lavaud2021}. The potential
minimum, $F(h^*) = 0$, is at $h^* = \ell_D\ln(B\ell_B/\ell_D) =
0.169\,\mu$m, where $\Phi = D_\perp' = 0.072\,\mu$m/s
$= 4.3\,\mu$m/min.

\emph{Moment route.} Chebyshev polynomials in $\ln h$ on $0.04 < h <
3\,\mu$m, degree $8$ for the drift and force and $16$ for the score;
$\nu$ evaluated on $0.1 < h < 1.5\,\mu$m ($91\%$ of the frames);
$\Dbar = 0.0325\,\mu$m$^2$/s; drift-corrected stencil for $\D_t$ (no
localization noise detected by the engine); standard errors from $10$
contiguous time blocks. Results: $\nu = 0.74\pm0.15$ (floor $0.084$,
detected); in relative $L^2$ error over the data points of the $\nu$
window, as in Table~\ref{tab:mr}, $\hat\sigma$ reproduces $D_\perp'$ to
$19\%$ and $\hat{\mathbf F}$ reproduces $F$ to $42\%$, against $61\%$
for the naive $\hat\Phi/\Dbar$. With the default stencil of the code:
$\nu = 0.71\pm0.36$ (not detected at that error), $29\%$ and $43\%$.
Potentials (Fig.~3f) are $-\int\hat F\,\mathrm{d}h$, zeroed at $h =
1\,\mu$m; the naive potential deviates from $Be^{-h/\ell_D} +
h/\ell_B$ by $1.03\,k_{\rm B}T$ at $h=0.15\,\mu$m, $0.71$ at $0.3$ and
$0.23$ at $0.6\,\mu$m; on $0.15 < h < 1.2\,\mu$m the consistent one
deviates by at most $0.16\,k_{\rm B}T$ and $-\ln P$ by $0.24\,k_{\rm
B}T$.

\emph{Tag and track.} The ``whichever way the movie is played'' of the main text is tested as in Figs.~1c and~2d: conditioning on the frames
with $h$ within $\pm10$~nm of $h^*$ ($4451$ tags) and following
$\av{h(t_{\rm tag}+\tau) - h(t_{\rm tag})}$ gives a curve symmetric in
$\tau$ ($\chi^2/{\rm dof} = 0.57$ for the forward--backward
difference) whose initial slope, $0.05$--$0.07\pm0.02\,\mu$m/s forward
and $0.04\pm0.02\,\mu$m/s backward, is compatible with
$D_\perp'(h^*)$; the curve then relaxes towards the stationary mean,
$\av h = 0.60\,\mu$m, as any conditioned ensemble would. Because a
genuine force also produces a symmetric relaxation, the test isolates
the spurious drift only through its initial slope at the point where
the force vanishes.

\section*{S11. Code and data}
Every figure and number is regenerated by one script per figure in the
reproducibility package (\texttt{repro/}), built on SFI
version~2.1.0~\cite{sfi_zenodo}, which includes the multiplicative-noise
module used for the moment route. The experimental trajectory of Sec.~S10 is from the
public repository of Ref.~\cite{lavaud2021}; the monolayer videos of Sec.~S9 are
from the public repository of Ref.~\cite{king2025}. The package will be made public with the published version of this article.

\end{document}